\documentclass{aa}
\usepackage{graphicx}
\begin{document}
\title{Detection of  Knots and Jets  in 
  IRAS 06061+2151}
\author{B.G. Anandarao \inst1  
            A. Chakraborty \inst2  
            D.K. Ojha \inst3    
	\and L. Testi\inst4}
\offprints{B.G. Anandarao, \email{anand@prl.ernet.in}}
	
\institute{Physical Research
           Laboratory, Ahmedabad-390009, India 
\and Department of Astronomy \& Astrophysics,
   525 Davey Laboratory, Pennsylvania State University, 
    University Park, PA 16802, USA 
\and Tata Institute of Fundamental Research,
Homi Bhabha Road, Mumbai-400005, India  
\and Osservatorio Astrofisico di Arcetri, INAF, Largo E. Fermi 5, 
     50125, Firenze, Italy \\ }
\date{}
\titlerunning{Knots and Jets in IRAS 06061+2151}
\authorrunning{Anandarao et al.}

\abstract{We report detection of a Young Stellar Object with an evidence for an
outflow in the form of knots  in the molecular hydrogen emission line
(2.121$\mu$m) towards the massive star forming region  IRAS 06061+2151. 
Near-infrared  images reveal IRAS 06061+2151 to be a cluster of at least five
sources, four of which seem to be early B type young stellar objects,   in a
region of  12 arcsecs surrounded by a nebulosity. The presence of  the knots
that are probably similar to the HH objects in the optical  wavelengths,
suggests emerging jets from one of the  cluster members. These jets appear
to excite     a pair of knot-like objects (Knot-NW and Knot-SE) and extend over
a projected size of 0.5pc. The driving source for the jets is traced back to a
member of the cluster  whose position in the H-Ks/J-H color-color diagram
indicates that it is a  Class I type pre-mainsequence star.   We also obtained
K band spectra of the brightest source in the cluster and of the nearby nebular
matter.  The spectra show molecular hydrogen emission lines but do not show  Br
$\gamma$ line (2.167$\mu$m).  These spectra suggest that  the excitation of 
the molecular hydrogen lines is probably  due to a mild shock.    

}

\maketitle

\section {Introduction}

IRAS 06061+2151 is one of the most luminous IRAS sources believed to be
associated with  the galactic massive star forming region Gemini OB1 which is
at a distance of 2 kpc (Carpenter, Snell \& Schloerb 1995a). The object 
figures in the radio survey of massive star
forming regions done by  Kurtz, Churchwell \& Wood (1994; hereafter KCW94). 
The CO survey of Shepherd \& Churchwell (1996) categorized this object as 
having a moderate velocity gas flow (full width = 20.8 km/s and fwhm = 4 km/s).
This Ultra Compact HII(UCHII) region was previously studied   by Carpenter,
Snell \& Schloerb (1995a) and very recently in a near-infrared (NIR) survey on
UCHII regions by Hanson, Luhman  \& Rieke (2002; hereafter HLR02). The
extensive CS survey and NIR observations by Carpenter, Snell \& Schloerb 
(1995a,b) of 58 IRAS sources in the Gemini OB1 molecular cloud complex 
suggested that more luminous IRAS sources  tend to be associated with more
massive cores. Based on their study, these authors  proposed a qualitative
model for the formation of massive dense cores, primarily   through the
external compression that sweeps shells of molecular gas. 

The mass outflows in the form of jets from young stellar objects (YSOs)  are
the important signatures of star formation in  molecular clouds (Hartmann 1998
and references therein). The jets in turn cause Herbig-Haro (HH)  
objects seen in
optical emission lines notably [SII] and H$\alpha$ by shocking/interacting 
with the ambient interstellar medium.  Such manifestations are better revealed
in the NIR due to the lesser extinction that it  suffers compared to  the
visible region. In the recent past several discoveries have been made of the
molecular hydrogen outflows; both in situations where there exist optical 
counterparts in the form of HH objects 
(eg., Schwartz etal 1988; Davis, Eisloffel \& Ray, 1994; 
Eisloffel, Smith \& Davis 2000) and where there 
are no optical counterparts (eg., Herbst, Beckwith \& Robberto 1997). However,  
the outflows detected in the case of massive proto-stars are rare due to the 
fact that the outflows are short-lived compared to those in the low mass 
proto-stars, though it is believed that stars of all masses might undergo
these phenomena.  

Therefore, luminous IRAS sources such as the IRAS 06061+2151 are   the
appropriate candidates for further study in the NIR. Our motivation  is to look
for new outflows associated  with massive protostars and to determine their
evolutionary stages (such as Class I, II and III), using color-color diagrams
(Lada \& Adams 1992; Lada, Young \& Greene  1993;  Gomez, Kenyon \& Hartmann
1994).  

In this paper, we report detection of infrared counterparts of HH objects
presumably caused by  jets from a class I type source  towards IRAS
06061+2151.  We study the nature of the cluster of sources using JHK
photometry, narrowband images in molecular hydrogen (2.12$\mu$m) and
Br${\gamma}$ lines and medium resolution K band spectroscopy.  The data were
obtained by new observations as well as from the 2MASS archive.  In section 2
we present the details of observations and data reduction  procedures, section
3 deals with results and discussion and we summarize our conclusions  in
section 4.

\section {Observations and Data Reduction}

\subsection {Observations from Mt. Abu}

J(1.253$\mu$m/0.296$\mu$m), H(1.659/0.288), K'(2.120/0.360) photometric
observations of IRAS 06061+2151  were done from the  1.2 meter PRL Infrared
Telescope, Mt. Abu, India,  using a 256$\times$256 HgCdTe  array (NICMOS-3,
made by Infrared Laboratories, Arizona, USA). The instrument  is described in
Anandarao (1998) and Nandakumar (1999). The plate scale is
0.5$^{\prime\prime}$/pixel (corresponding to a field of view of 2$^{\prime}$).
The night of observation (26 Jan 2001) was photometric with a seeing of
0.9-1.0$^{\prime\prime}$. 

Individual object frames were of 30s of integration in J and H bands and 3s in
K' band. The data reduction was done using IRAF software tasks. All the images
were subjected to standard pipeline procedures like dark and sky subtraction 
and flat-fielding. The images were then co-added to obtain a final image in
each band (J, H and K') averaged over a total integration time of 300s. 
Further, in each band two such images were obtained. The images were then 
filtered using a median filter of  3$\times$3 pixels for removing noise at
sky-level, thus making them suitable for photometric analysis. We used DAOPHOT
(Stetson 1987) task and followed the photometric procedure described by
Chakraborty etal. (2000).  We estimate the overall photometric error to be 
$\pm$0.08 mag.  The faintest stars recorded were found to be of 16.66, 15.30
and  14.10 magnitudes in J, H and K' bands respectively.

\subsection {2MASS data}

We extracted sources from the 2MASS All-Sky Point Source Catalogue(PSC)  which
were within 2$^{\prime}$ radius around the IRAS source.  Thus this region
excludes  the cluster G188.80+01.07 (see Bica, Dutra \& Barbuy 2003). The data
were downloaded from the 2MASS Homepage available for the Astronomical
community. The 2MASS point source catalogue consists of 
J(1.24$\mu$m/0.33$\mu$m),  H(1.66/0.39) and Ks(2.16/0.42) magnitudes of
stars.   The PSC detection limits for a S/N=10 are J=15.8, H=15.1 and Ks=14.3. 

The 2MASS data are used to produce the color-color diagram to study the nature
of sources. We use the Mt. Abu  photometry for the five bright cluster
candidate YSOs towards IRAS 06061+2151(see Sec. 3.1).   It is to be noted that
the Mt.Abu magnitudes match reasonably well with the 2MASS  data as shown in
Table 1 (except for Stars 1 and 5), inspite of the differences in the
passbands. 

\subsection {Observations from TNG, La Palma}

We obtained spectra (R=1000, 1.4 to 2.4 $\mu$m) of one of the bright sources
(Star no. 1) in the cluster of IRAS 06061+2151 using the Near Infrared Camera 
Spectrometer (NICS) attached to the 3.58 m Telescopio Nazionale Galileo (TNG) at
La Palma.  NICS has a HgCdTe Hawaii 1024x1024 array and it was used under LF
configuration with a plate scale of 0.25$^{\prime\prime}$(for instrument details, see
Baffa etal. 2001).  The medium
resolution spectra were obtained using the HK grism with a slit width of
0.5$^{\prime\prime}$. The Position Angle (PA) of  the slit was 270 degrees i.e.,
in the East-West direction. Two dithered exposures of 60 secs each were
obtained on the Star no. 1.  A hot star (A0) spectra were also taken for
telluric calibration. Following a similar procedure, spectra were obtained
in a region 5$^{\prime\prime}$ west of Star no. 1, in order to sample 
the nebular regions. 

Apart from the spectra, we also obtained images of the source in two
narrowband  filters centered around 2.12$\mu$m for the H$_2$ v=1-0 S(1) line
and 2.17$\mu$m for the  Br$\gamma$ line. A total of five 60 seconds dithered
exposures (dithered by more than 3 arcmins) were obtained in both the
narrowband filters.  The seeing was 0.8-0.9 arcsec during observations. 

The TNG images were analysed in a similar process as those from Mt. Abu. The
spectra  were reduced using standard IRAF tasks like APALL, TELLURIC, and
normalized with  respect to the respective continua. The wavelength calibration
was done using  the sky OH lines using the list of lines available with IRAF.
We present here  the K band spactra for which the signal to noise ratio  is
better than 5. This spectral region contains some of the most important
features  that we are looking for, such as the Br$\gamma$ and H$_2$ vibrational
lines.  

\section {Results and Discussion}

\subsection {Photometry}

Fig 1 shows the K' band image of the central region of the object taken at Mt
Abu. The five prominent stellar sources (S1,2,3,4 and 5) are  marked on the
figure and their   magnitudes are given in Table 1. The data for the S4 (all
three bands) and S5 (only in K' band) are from  the Mt. Abu observations.  S4
is detected in H and Ks bands but not in J band; while S5 is detected  only in
Ks band in 2MASS.  The Mt Abu K' magnitude of S5 is 12.9 while in J and H bands
the source was below the  detection limit. HLR02 obtained an H band image of
the central region of this object which does not show S5; thus lending support 
our result. It appears then that  this source has a rather steep spectral
energy distribution in the NIR and beyond.   Hence the importance of the NIR
study at high angular  resolution (atleast seeing-limited) of the massive star
forming regions. 

The IRAS flux densities for 06061+2151 are 12.0, 144.8, 865.6 and 1130 Jy in
the 12, 25, 60 and 100 $\mu$m bands respectively. From these values, 
Carpenter, Snell \& Schloerb (1995b) estimated the total far-infrared (FIR)
luminosity to be 4000 L$_{\odot}$. Based on the total FIR luminosity, Ghosh
etal. (2000) suggested that the core may contain a single B0-B0.5 type hot
star, though they could not resolve the cluster due to the large field of view
(1.5 arcmin). Our NIR image (see Figure 1), however, reveals that IRAS 
06061+2151 consists of a cluster of at least five stars. 

Figure 2 shows J-H/H-Ks color-color diagram of sources from the 2MASS data and
Mt. Abu observations (S1-S4). The 2MASS-PSC data covers a diameter of 4 arcmins
around the co-ordinate of the IRAS source. The total number of sources
extracted were 57, after removing all the "null" entries.  The solid curve is
the locus of points corresponding to unreddened main sequence stars and
giants (Bessell \& Brett 1988).  The two parallel lines (dashed) are the
reddening vectors drawn upto the visual extinction of Av = 30; together these
parallel  lines show the region in which the main-sequence stars occur at
various  extinctions (Hunter etal. 1995). Sources on the right side of the
reddening strip ($\sim$ 17) are primarily stars 
with intrinsic color excess which
include the S1-S4 stars of the cluster.   We note here that we allowed an
uncertainity of 0.1 mag in H-K color on either side of the strip to identify
the left and right side of the reddening strip. Also shown in the figure is the
locus for the CTTS (from  Meyer, Calvet \& Hillenbrand 1997) and the 
corresponding reddening vector. 

The stars on the right side are mostly Young Stellar Objects (YSOs) (Lada \&
Adams  1992; Lada, Young \& Greene 1993; Gomez, Kenyon \& Hartmann 1994).  In
general, sources whose H-K color is  $\ge$1.5 and J-H $\ge$1.0 can be
considered as Class I type YSOs or protostars in  the early stages of their
evolution (Strom, Margulis \& Strom 1989; Kenyon   et al. 1993; Lada  \& Adams
1992). From Fig 2, we find that out of the total of 17 stars falling clearly in
the region  on right side of the reddening vector, 3-4 are clearly beyond the
CTTS locus to its right.  These stars that include S4 and maybe S2 can be
classified as possible Class I  candidates, if we account the uncertainty of
0.1 mag in colors.   The rest  numbering about 13-14 stars can be classified as
possible Class II - III.  Thus, the region around IRAS 06061+2151 shows a
concentration of  YSOs.   Indeed, NIR studies by Carpenter, Snell \& Schloerb
(1995b) predicted a similar trend  towards IRAS 06061+2151.

Figure 3 shows H-Ks/Ks color-magnitude diagram. Like in figures 1 and 2 the
stellar sources (S1,2,3 and 4) are marked. The solid line represents unreddened
main-sequence stars (Koornneef 1983; Bessell \& Brett 1988)  and the dashed line
is the reddening vector of Av = 30 mag. S1  shows a high extinction of Av = 28.
The other sources S2, S3 and S4 also show high extinctions of Av = 19, 26 and
36 respectively. Based on the photometry in Table 1,  using Rieke \&
Lebofski (1985) extinction law and a  distance modulus = 11.5 mag, the expected 
absolute K magnitudes of  the sources S1, 2, 3 and 4 are Mk = -3.6, -2.4, -2.8
and -3.0 respectively. Source S5 which is only observed in the K' band perhaps
suffers extinction greater than Av = 40 mag, if we consider that it is fainter
than the detection limits of the Mt. Abu observations in J and H bands.

If the above arguments are true,   then the sources have to be much more
massive than T Tauri stars.  Following the Palla \& Stahler (1999) evolutionary
tracks, stars less massive than 3.5 M$_{\odot}$ (at any age)  are never more
luminous than Mk = -1.67. Assuming that more massive  stars actually burn
hydrogen while accreting, they are expected to have no pre-mainsequence
evolution  (as the authors above advocate). Using the zero-age main sequence
magnitude/color relations Mk = -3.6 implies a spectral type between B0 and B1 and
Mk = -2.4 corresponds to B2. We have considered IR colors from Bessell \& Brett
(1988) for our calculations.  In all these arguments, we have neglected
infrared excess and accretion luminosity. 

For the source S1, being the most massive,  the accretion luminosity could
contribute substantially to the luminosity only if the accretion  rate is very
high. This is possible, but implies an amount of circumstellar material that
would be so high to produce a column density much higher than Av = 28 mag. For
other sources S2, 3 and 4 the uncertainities in the infrared excess will apply,
however, the absolute magnitude-luminosity relation does give us a lower limit
for the stellar masses.   

The important finding of the current photometric study is Star S5 whose J and H
magnitudes are fainter than 15.2 and 16.4 magnitudes (our detection limits)
while the K' magnitude is  12.9 (2MASS data gives 13.52).  The absence of S5 in
the H band image of HLR02 confirms this. In any case a deeper NIR survey
preferably at high  spatial resolution may show exactly how steep is the NIR
energy distribution for S5 and hence its exact nature. The difference in the K
magnitudes from Mt Abu and  2MASS is 0.62 and is larger than the observational
uncertainties. It is possible  that the star may be undergoing a variability
that is akin to FU Orionis. Such a substantial difference is also noticed (Table
1)  in the case of K band for  S1.   

\subsection{Molecular Hydrogen Knots in IRAS 06061+2151}

Figures 4 and 5 show narrow band images of the region in H$_2$ (2.121$\mu$m,
line + continuum) and  Br$\gamma$ (2.167$\mu$m, line + continuum)
respectively.  The circle marks  two knot-like diffuse structures, one in the
north-west and another in the south-east  of the centre of the cluster in the
H$_2$ image. These two prominent diffuse emission features are  not seen in the
Br$\gamma$ image. We may safely conclude then that the narrow band image  at
2.167 $\mu$m (Fig 5) represents the background continuum radiation.  We
estimated the integrated  brightness of the knots using  relative photometry
with respect to the Ks magnitude of S1 and S2.   A large  aperture of 20
arcsecs was used for the knots. We find that in the  H$_2$ image, the western
knot is 1.6 mag and the eastern knot 2.5 mag fainter  than S1. A similar
estimation of brightness of faint stars seen in both the H$_2$ and Br$\gamma$
images showed that they are 4 mag fainter. We also find that these
knot-like structures should be at least 4.5 mag fainter than S1 in the
Br$\gamma$ filter since they are not detected in the image. 

We thus establish that the two newly identified structures are primarily due to
H$_2$ emission. The H$_2$ $v$ = 1-0 S(1) 2.122$\mu$m line emission has been
proven to be an excellent tracer in dense molecular cloud regions for moderate
shocks in the range of 10 to 30 km/s (Shull \& Beckwith 1982;  Draine, Roberge,
\& Dalgarno 1983; Sternberg \& Dalgarno 1989) and are found to trace the
termination of CO outflows and optically seen HH objects (see  eg.
BACHILLER 1996; Hartmann 1998 and references therein). 
It is possible that due to this moderate
nature of the shock the ionization of hydrogen could not take place. Therefore,
we propose that this pair of  knot-like structures is probably tracing a shock
caused in the surrounding  interstellar medium by protostellar jets.  We call
them Knot-NW and Knot-SE respectively.  These knot-like structures may be
considered as infrared counterparts  of HH Objects that are often
detected in [SII] and H $\alpha$ images. From  the H$_2$ image, we determine
the co-ordinates ($\alpha$(2000), $\delta$(2000)) of the Knot-NW and Knot-SE as
(06h09m05.2s, +21$^{\circ}$50$^{\prime}$56$^{\prime\prime}$) and (06h09m08.5s,
+21$^{\circ}$50$^{\prime}$32$^{\prime\prime}$) respectively. Judging from the
brightness of the knots, we assume that Knot-NW suffers lesser extinction  and
hence is tracing a blue jet and the Knot-SE the red one. Alternatively, the
difference  in brightness could also be due to the differences in physical
parameters such as density  of the two regions of interstellar medium.       

There are two sources, namely S1 and S4 lying in the line of the jets.  From
the color-color diagram (Fig 2), it appears that S1 is likely a massive  
member of the cluster and a Class II source,  while S4 is less massive but 
clearly fits  in the scenario of a  Class I protostar.  So it is tempting to
suggest that S4  could be more likely  the source of the jets.  It is also
situated nearly in the middle of the projected distance between the two knots. 
The projected length of the jet is estimated  to be about 0.5 pc.

\subsection{Medium Resolution TNG K Band Spectra} 

Figure 6 shows in the upper panel normalized K band spectra,  taken at TNG, 
towards the star S1 and  in the lower panel  the nebular spectra obtained in a
region 5 arcsec  west of S1 and towards S4 (close to the path of the NW jet;
see Figure 4). Our main  concern is the H$_2$ line ratio of 2-1 S(1)/1-0 S(1)
(2.247$\mu$m/2.122$\mu$m), which depends  critically on the type of excitation
mechanism namely fluorescence or shock heating (Shull \& Beckwith 1982; 
Draine, Roberge \& Dalgarno 1983; Black \& van Dishoeck 1987; Sternberg \& 
Dalgarno 1989; Draine \& Bertoldi 1996; Luhman, Engelbracht \& Luhman 1998; 
Walmsley etal 2000; line ratios compiled in Glass 2000). 
Since the 2.247$\mu$m line is very weak and buried in the base-level noise, 
we put an upper limit for the
ratio at 0.1 and 0.05 for the nebular spectra and the Star 1 respectively  
by taking the base line as 1.0$\pm$0.1 for nebular spectra and 1.0$\pm$0.05 
for the Star 1 spectra. 
therefore, these upper-limit ratios strongly suggest that the excitation 
could be due to a shock rather than fluorescence where one would expect the 
ratio to be $\geq$ 0.5 (eg., Black \& van Dishoeck 1987; 
Draine \& Bertoldi 1996). One can notice the 2.223$\mu$m
line of H$_2$ 1-0S(0) appearing quite prominently in both the spectra with 
the ratio 1-0S(0)/1-0S(1) equal to 0.55 for nebula and 0.35 for star 1. 
these ratios 
are rather high for collisional excitation. but at an equilibrium 
temperature of 1000 K, the ratio of 1-0S(0)/1-0S(1) 
rises (to $\sim$ 0.3), while the ratio of 2-1S(1)/1-0S(1) falls   
to insignificance (see table 2 in Black \& van Dishoeck 1987).  
The fluorescence excitation populates the vibrational levels v$\geq$2 much 
more than in the thermal excitation; this does not seem to be the case here. 
We therefore infer from the spectra that the lines are likely thermally
excited.      
The ratios for S1 are in reasonably good agreement with the 
ratios obtained by HLR02. Prominently missing in the spectra is the 
Br${\gamma}$ emission line at 2.167$\mu$m  which could be due to the fact 
that the line if present is very weak.  The absence of Br $\gamma$ is also 
reported by HLR02.
So it is clear from the presented spectra that the molecular hydrogen
excitation is caused by moderate shocks (with velocities less than about 30
km/s that can not cause hydrogen ionization as it requires greater than 60
km/s) which explains why Br${\gamma}$ emission line is absent or very weak.   

In Fig 6 we find the spectrum towards S1 (upper panel) to be significantly
weaker than the spectrum  on the nebular region towards S4 (lower panel). So if
S1 is indeed a class I protostar then we expect   the two spectra to be of more
or less comparable strengths (since S4 is a protostar).  This is a further
evidence that S1 is likely a massive and more evolved member of the cluster. 

HLR02 concluded that the radio source KCW 188.793+1.030 (R1) corresponds to the
source S1 while the spectra taken by them on the radio source KCW
188.796+1.030 (R2) do not seem to  have a counterpart in their H band image. But
our K' image taken at Mt Abu as well as the TNG narrow band images show that
the S5 source does line up with HLR02 spectra on R2. The line ratios reported
by HLR02 on R2 (S5)  are nearly similar to those on R1 (S1).  The absence of Br
$\gamma$ in the two spectra presented by HLR02 and the TNG spectra  shows that
in general the region may be pervaded by a mild shock with velocities not
greater than  30km/s. 

\section {Conclusion}

IRAS 06061+2151 appears to be a cluster of at least 5 bright sources  four of
which seem to be early B type YSOs. From the J-H/H-Ks color-color diagram we
found that one of them, S1, could be the most massive  member of the cluster
rather than a background source and  S4 is a Class I type. S3 seems to be a
Class II source while  S2 is likely to be Class I type. The IRAS source is also
associated  with a nebulosity. As many as 3-4 Class I (including S4 and maybe
S2)  and 13-14 Class II/III sources (including S1 and S3)  were  found towards
the region of IRAS 06061+2151 from the color-color diagram which confirms the
earlier studies that the region contains large number of YSOs. A new source S5
appears to have a very steep infrared energy distribution and is likely  a
massive protostar as its colors (lower limits) suggest a heavily embedded
source.

From the narrow band images in H$_2$ and Br${\gamma}$, two oppositely directed 
knot-like structures (Knot-NW and Knot-SE) have been detected. The knots are
likely generated by jets from the  protostellar source S4 and are the infrared
counterparts of the classical HH objects in optical region. From the ratios of
H$_2$ lines, we conclude that the region is probably pervaded by a mild shock
or a photodissociation region that can excite these lines  but not strong
enough to excite the HI lines.

\begin{acknowledgements}

We thank the referee Dr. M. Hoogerheijde for valuable suggestions 
for improving the original version.   
It is a pleasure to thank  the staff at Mt. Abu IR Telescope and the TNG, La
Palma facility   for excellent support during observations. This work is based
in part on the observations  made with the Italian Telescopio Nazionale Galileo
at La Palma operated by the Centro  Galileo Galilei of CNAA at the Observatorio
del Roque de los Muchachos of the  IAC, Canary Islands, Spain.  This
publication makes use of data products from  the 2MASS, which is a joint
project of the University of Massachusetts and the Infrared Processing and
Analysis Center, funded by the NASA and the NSF. This research has made use of
the SIMBAD database, operated at CDS, Strasbourg, France.    

\end{acknowledgements}

\newpage

\begin{table*}
\caption[]{A comparison of stellar magnitudes from 2MASS data and Mt. Abu
images.}
\[
\begin{tabular}{|c|c|c|c|c|c|c|c|}
\hline
{\it No.}&{\it Co-ordinates (J2000)}&
 \multicolumn{3}{c|}{2MASS} & 
  \multicolumn{3}{c|}{MtAbu}  \\ \cline{3-8}
&{\it RA-DEC}&{\it J}&{\it H}&{\it Ks}&{\it J}&{\it H}&{\it K'}\\
\hline
1&06h09m07.2s +21$^{\circ}$50$^{\prime}$40$^{\prime\prime}$&
15.46&12.87&11.22&15.54&12.68&10.85\\
2&06h09m07.1s +21$^{\circ}$50$^{\prime}$34$^{\prime\prime}$&
13.67&12.14&11.20&13.65&12.24&11.11\\
3&06h09m07.5s +21$^{\circ}$50$^{\prime}$32$^{\prime\prime}$&
15.31&12.94&11.55&15.44&13.05&11.45\\
4&06h09m06.9s +21$^{\circ}$50$^{\prime}$43$^{\prime\prime}$&
--&14.34&12.16&16.66&14.48&12.29\\
5&06h09m07.1s +21$^{\circ}$50$^{\prime}$32$^{\prime\prime}$&
--&--&13.52&--&--&12.90\\
\hline
\hline
\end{tabular}
\]
\end{table*} 

\newpage

\begin{figure}
 \resizebox{\hsize}{!}{\includegraphics{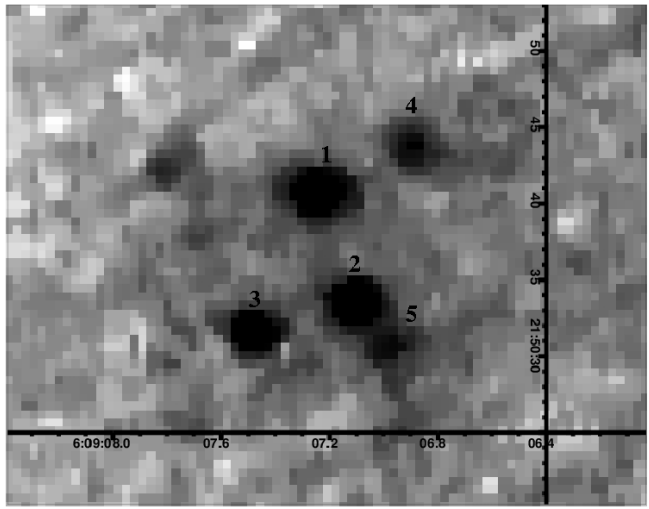}}
  \caption{K' band image of IRAS 06061+2151 observed from Mt Abu. The
numbers represent the cluster of sources around IRAS 06061+2151 region
The abscissa and ordinate are in J2000.0 epoch.}
\label{a}
\end{figure}

\newpage

\begin{figure}
 \resizebox{\hsize}{!}{\includegraphics{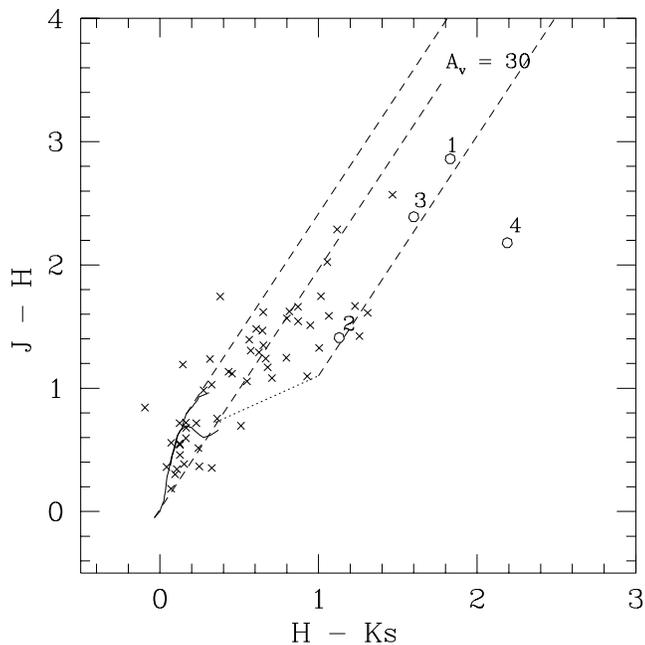}}
  \caption{Color-color diagram of the sources extracted from the 2MASS-PSC 
   data. The numbers represent the cluster of sources around IRAS 06061+2151
   region (see text and 
   Table 1 for details).
The solid line represents unreddened main-sequence stars and the giants 
(Bessell \& Brett 1988). The dashed lines
are parallel to the reddening vector with magnitude of Av=30. 
 The dotted line is the locus of CTTS (Meyer etal. 1997).}
\label{a}
\end{figure}

\newpage

\begin{figure}
 \resizebox{\hsize}{!}{\includegraphics{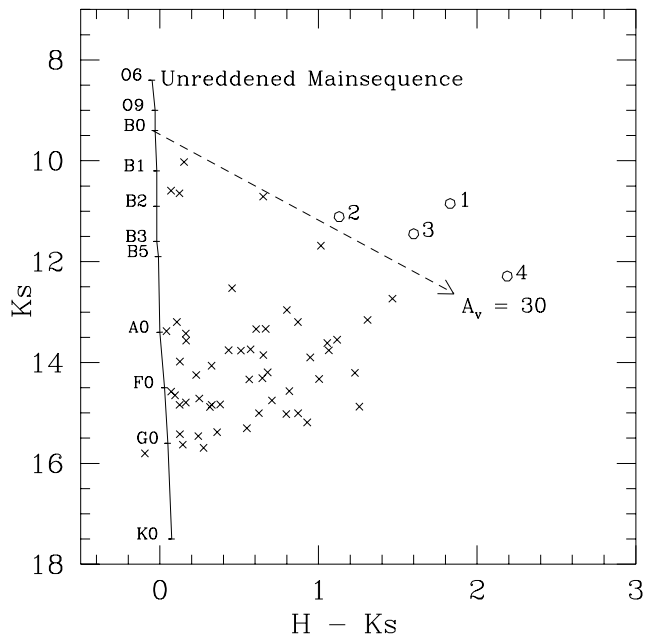}}
  \caption{Color-Magnitude diagram of the sources extracted from the 2MASS-PSC 
  data. 
  The numbers represent the cluster of sources around IRAS 06061+2151 region. 
The solid line represents unreddened main-sequence stars (Koornneef 1983; 
 Bessell \& Brett 1988)
 and the dashed line
is the reddening vector with magnitude of Av=30.}
\label{a}
\end{figure}

\newpage

\begin{figure}
 \resizebox{\hsize}{!}{\includegraphics{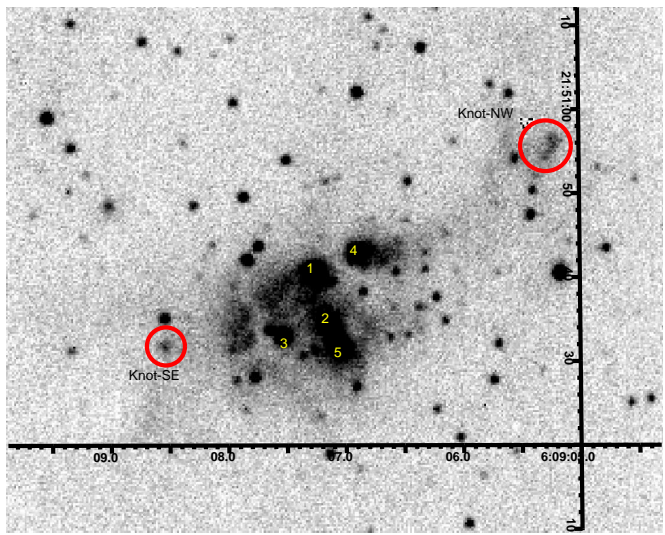}}
\caption{H$_2$ line + continuum (2.122$\mu$m) image observed from TNG. 
 The circles show the Knot-like objects Knot-NW and Knot-SE. 
 The numbers 1,2,3,4 and 5 mark the positions of the stars as
numbered in Table 1. North is up and East is to the left.
 The abscissa and ordinate are in J2000.0 epoch.}
\label{a}
\end{figure}

\newpage

\begin{figure}
 \resizebox{\hsize}{!}{\includegraphics{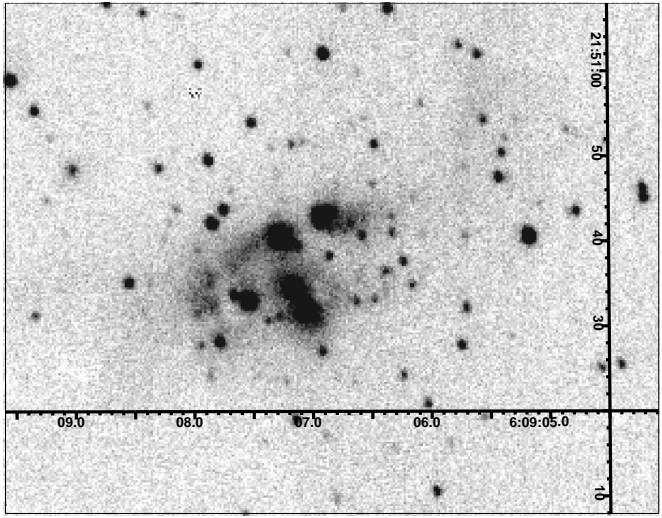}}
 \caption{Br${\gamma}$ line + continuum (2.167$\mu$m) image observed from TNG.
  North is up and East is to the left.
   The abscissa and ordinate are in J2000.0 epoch.}
\label{a}
\end{figure}

\newpage

\begin{figure}
 \resizebox{\hsize}{!}{\includegraphics{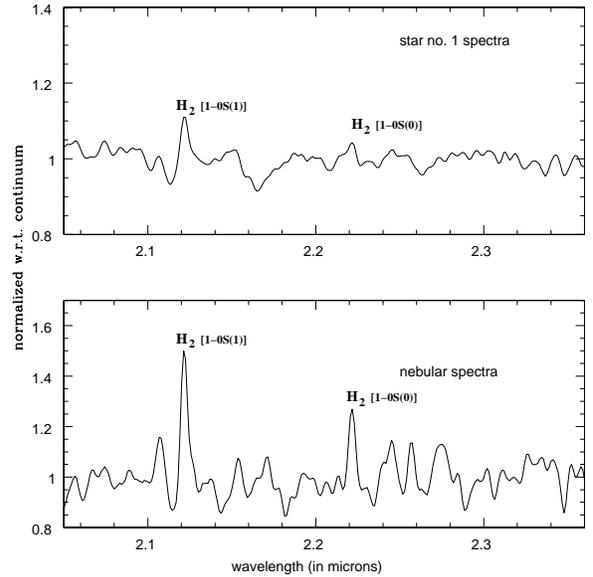}}
 \caption{Spectra towards Star 1(top) and that of nebula (bottom) 
  which is 5 arcsec west of Star 1 towards Star 4. Both the spectra are 
  normalized with respect to its continuum.}
\label{a}
\end{figure}

\end{document}